\documentclass{Rpd}
  \usepackage[T1]{fontenc}
  \usepackage[utf8]{inputenc}
  \usepackage{siunitx}
\begin{document}



\title[Cellular burdens and biological effects of radon progenies]{Cellular burdens and biological effects on tissue level caused by inhaled radon progenies}
\author[Madas \textit{et~al}]
{
  Balázs G. Madas\,$^{\rm a}$\textsuperscript{,}\footnote{Corresponding author: balazs.madas@energia.mta.hu}~,
  Imre Balásházy\,$^{\rm a,b}$,
  Árpád Farkas\,$^{\rm a}$,
  István Szőke\,$^{\rm a}$
}
\address{$^{\rm a}$Health and Environmental Physics Department, Hungarian Academy of Sciences KFKI Atomic Energy Research Institute, Konkoly-Thege Miklós út 29-33, H-1121 Budapest, Hungary, $^{\rm b}$Aerohealth Ltd., Csillag sétány 7, 2090 Remeteszőlős, Hungary}

\date{This is a pre-copy-editing, author-produced PDF of an article accepted for publication in Radiation Protection Dosimetry following peer review. The definitive publisher-authenticated version (Madas, B. G., I. Balásházy, Á. Farkas, and I. Szőke. "Cellular burdens and biological effects on tissue level caused by inhaled radon progenies." Radiation Protection Dosimetry 143 (2011): 253-257. doi:10.1093/rpd/ncq522.) is available online at: {http://rpd.oxfordjournals.org/cgi/reprint/ncq522?ijkey=vyzsY5jowZYZjT0\&keytype=ref}.}

\begin{abstract}
In the case of radon exposure, the spatial distribution of deposited radioactive particles is highly inhomogeneous in the central airways. The objective of this research is to investigate the consequences of this heterogeneity regarding cellular burdens in the bronchial epithelium and to study the possible biological effects on tissue level. Applying a computational fluid dynamics program, the deposition distribution of inhaled radon daughters has been determined in a bronchial airway model for 23 minutes of work in the New Mexico uranium mine corresponding to 0.0129 WLM exposure. A numerical epithelium model based on experimental data has been utilized in order to quantify cellular hits and doses. Finally, a carcinogenesis model considering cell death induced cell cycle shortening has been applied to assess the biological responses. Computations present, that cellular dose may reach 1.5 Gy, which is several orders of magnitude higher than tissue dose. The results are in agreement with the histological finding that the uneven deposition distribution of radon progenies may lead to inhomogeneous spatial distribution of tumours in the bronchial airways. In addition, on macroscopic level, the relationship between cancer risk and radiation burden seems to be non-linear.
\end{abstract}

\maketitle

\section{Introduction}

Radon is considered to be the second most important cause of lung cancer after smoking \cite{EPA01}. Inhaled radon-daughters deposit in the respiratory system and most of the alpha-particles emitted by them hit the surrounding epithelial cells. Previous studies proved that the spatial distribution of deposited radioactive particles is highly inhomogeneous in the central airways \cite{Bal01}, \cite{Far01}, where radon induced malignancies predominate \cite{Sch01}. Therefore, it is quite important to determine what consequences this heterogeneity has and how cancer risk depends on the amount of inhaled radon progenies. The objective of this research is to investigate these questions applying numerical modeling techniques. 

\section{Methods}

The elaborated computational modeling efforts consist of three major steps. Firstly, a 3D numerical epithelium model and a suitable microdosimetric model have been developed for the quantification of cellular hit numbers and doses. The next step is preparing a radiation biological carcinogenesis model, which connects biological responses to the computed microdosimetric data. Finally, this model can be applied for a given deposition distribution computed by computational fluid and particle dynamics approaches.

Brick shaped epithelium model has been elaborated taking into account the experimental data about the average volumes, frequencies \cite{Mer01} and depth distributions \cite{Mer02} of the six epithelial cell types in the bronchial epithelium.
The height of the fragments is 57.8~\SI{}{\micro\metre},
while its area is equal to the area of individual numerical grid elements to be modeled constructing the airway geometry. The cells in the model are brick-shaped and placed in three layers: basal cells can be found at bottom, goblet and ciliated cells are placed in the upper region, other secretory and preciliated cells are located either in the upper or in the middle layer, while intermediate cells are situated either in the middle or in the bottom part of the epithelium model. There is no gap between the cells. The model of the epithelium is covered with a 5-micrometer-thick mucus layer.

In the microdosimetric model, the dose-contributions of beta- and gamma-radiation are neglected, i.e. only the alpha-particles are considered. The path of alpha particles is supposed to be straight. The deposited energy along a given track length and the range is determined by the utilization of  \textquotedblleft The Stopping and Range of Ions in Matter\textquotedblright  software \cite{SRI01}. The energy lost in the airway lumen is considered by the application of the model presented in a previous work \cite{Szo01}. Calculating the intersection of brick-shaped cells and the tracks of alpha-particles, cellular hit numbers can be determined. In addition, knowing the alpha-energies at the boundaries of the cells, cellular absorbed energies and doses can be computed.

For the assessment of biological effects, a revised version of the so-called Initiation-Promotion model \cite{Tru01} has been applied. This model supposes that carcinogenesis consists of two stochastic steps. The first one is called initiation, which means transformation in this model. Initiation frequency ($p_\mathit{ini}$) is directly proportional to cellular absorbed dose ($D$):
\begin{equation} 
  p_{\mathit{ini}} \propto D\label{eq:01}
\end{equation}

According to the model, the second and final stochastic step towards cancer is promotion, which means the division of an initiated cell. In the bronchial epithelium, basal and goblet cells are thaught to be able to divide \cite{ICR01}, therefore initiation and promotion frequencies are computed only for these two cell types.

Supposing that cell death forces other cells to divide, promotion probability ($p_{pro}$) will be directly proportional to mitotic rate, which consists of two terms in the model. The normal mitotic rate is constant ($\lambda_{1}=1/30~ \mathrm{day}^{-1}$), while the forced mitotic rate is directly proportional to the number of cells died ($N_{d}$) and inversely proportional to the number of cells capable of dividing ($N_\mathit{cd}$):

\begin{equation}
   p_\mathit{pro} \propto \lambda_{1}+\lambda_{2} \cdot N_{d} \diagup N_\mathit{cd},\label{eq:02}
\end{equation}

where $\lambda_{2}$ (1 $\mathrm{day}^{-1}$) is constant.
 
The number of cells died in the individual grid elements is assessed by the expected value of cells killed utilizing the individual cell surviving probabilities ($p_\mathit{sv}$), which decreases exponentially with the increase of absorbed dose

\begin{equation}
p_\mathit{sv}=\exp(-\gamma \cdot D),\label{eq:03}
\end{equation} 

where $\gamma$ (1.67 $\mathrm{Gy}^{-1}$) is constant.

The probability that a given cell takes both stochastic steps are computed as the product of initiation frequency, promotion probability, and naturally, cell surviving probability, because only living cells are able to be initiated and promoted. Presuming that the value of this probability is low, the probability that at least one cell capable of dividing in a given surface element of the airway geometry takes the above two steps ($R$) can be expressed by (4):

\begin{equation}
R \propto \sum_{i}^{N_\mathit{cd}} D_{i} \cdot \exp(-\gamma \cdot D_{i}) \cdot (\lambda_{1}+\lambda_{2} \cdot N_{d}/N_\mathit{cd}).\label{eq:04}
\end{equation}

\begin{figure}
\centerline{\includegraphics{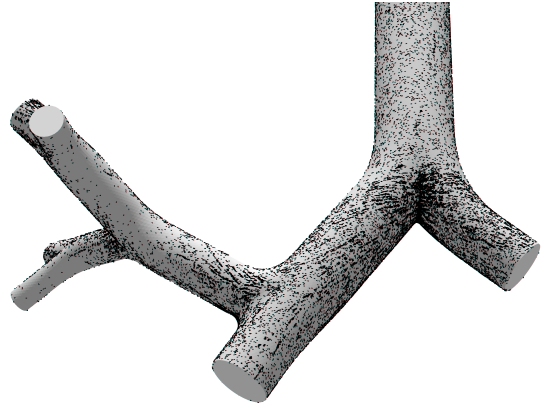}}
\caption{Deposition distribution of radon progenies in the geometry model. Black dots demonstrate the deposited particles. Due to impaction and other deposition mechanisms, particle free regions and areas with high particle accumulation occur in the surface of branches.}\label{fig:01}
\end{figure}

The deposition distribution is determined by com\-pu\-ta\-tional fluid and particle dynamics approaches. For the computations, the airway geometry composed of five bifurcation units was discretised by the construction of an inhomogeneous, unstructured numerical mesh \cite{Far01}. Air and particle flow describing equations were solved by finite volume methods. One way coupling has been assumed between the air and the radioactive particles. This presumes that particle trajectories are influenced by the airstreams, but the airflow is independent of the particles. Inhaled particles were tracked until they left the targeted airways or deposited on the airway walls. Since a viscous mucus layer is covering the bronchial airways, particle-wall impaction can be considered as inelastic. The multitude of these impaction points forms the deposition pattern. Primary deposition patterns can be modified by the clearance mechanisms, especially mucociliary clearance. However, clearance mechanisms were not considered in this study. The microdosimetric and biological quantities are computed for epithelium model fragments with the same size as the individual elements (grid cells) of the mesh. Then quantities characteristic for the whole geometry are calculated by the summation or averaging of quantities computed for the individual grid elements. For example, the right side of relation (4) has been computed for each grid cell of the geometry and summed up, so as to obtain a quantity directly proportional to cancer risk.

The deposition distribution in the bronchial airway model demonstrated in Figure \ref{fig:01} corresponds to 500 inhalations, 50 millions inhaled and 500~thousands deposited particles in the geometry during 23 minutes of work in the New Mexico uranium mine, which is equivalent to 0.0129~WLM. The geometry model consists of 162~thousands surface units (or grid elements), mean area of which is 62500~\SI{}{\micro\metre^2} with about 3300 epithelial cells. The biological responses in the different surface units are presumed to be independent.

\begin{figure}
\centerline{\includegraphics{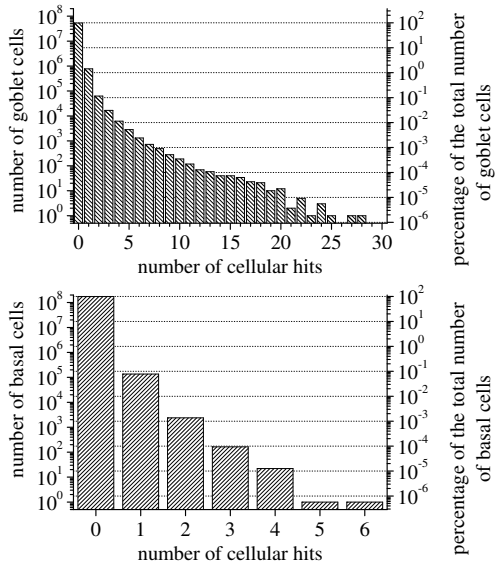}}
\caption{Cellular hit-distribution in case of goblet (upper panel) and basal cells (bottom panel) for the whole geometry.}\label{fig:02}
\end{figure}

\section{Results}

The cellular hit-distributions are plotted in Figure~\ref{fig:02} for basal and goblet cells. Most cells receive no hit, while some cells can suffer tens of hits. Obviously, basal cells are hit less frequently than goblet cells, since as compared to goblet cells, they are more deeply embedded in the epithelium. Importantly, the average hit number over all the cell types is 0.0079, more than three thousands times less than the maximum number of hits. In terms of percentages, 98.4\% of goblet cells receive no hit, 1.4 suffer single hits, while 0.2\% are hit multiple times. Regarding basal cells, 99.9\% percentage of them are not hit, however, there are cells receiving multiple hits, as well.

\begin{figure}
\centerline{\includegraphics{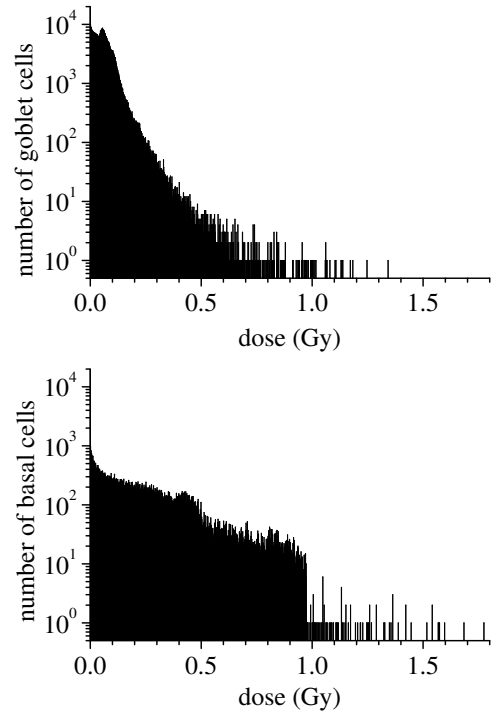}}
\caption{Cellular dose-distribution in case of goblet (upper panel) and basal cells (bottom panel). There is a cutoff at 0.97~Gy in case of basal cells, because doses above this value can originate solely from multiple hits, which are rare.}\label{fig:03}
\end{figure}

Figure \ref{fig:03} presents the distributions of absorbed doses at cellular level for cells hit. Doses are determined independently of the calculation of cellular hit numbers. The cellular absorbed doses vary in a wide range between 0~and 1.5~Gy. Interestingly, there are several cells, which absorb more than 1~Gy dose during the 23-minute-long exposure, while almost 99.3\% of all the cells receive no hits and therefore absorb no dose. The mean cellular dose of basal cells is 0.15~mGy, while that of goblet cells is 0.86~mGy for exposure characterized by 0.0129~WLM. The average cellular dose over all cells is approximately 0.5~mGy, what means that tissue dose is also of this order of magnitude, while some cells and some clusters of cells absorb thousand times higher doses.

\begin{figure}
\centerline{\includegraphics{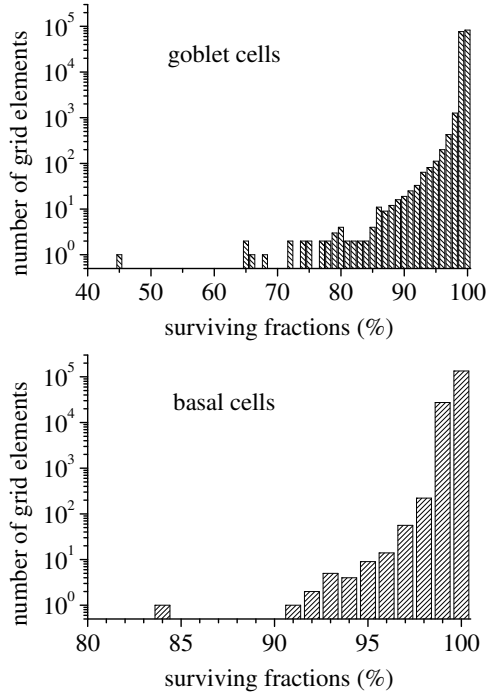}}
\caption{Number of grid elements characterized by a given surviving fraction of goblet (upper panel) and basal cells (bottom panel).}\label{fig:04}
\end{figure}

The fractions of surviving cells have also been computed in each grid element, so the distribution of surviving fractions is obtained. In Figure \ref{fig:04}, the number of surface units characterized by a given surviving fraction is depicted. The lowest surviving fraction in one grid element is 45\% in case of goblet cells and 84\% in case of basal cells. It is worth to mention that these low surviving ratios in a grid element are the result of only 23 minutes of radon exposure corresponding to 0.0129~WLM. On the other hand, surviving ratios lower than 99\% are quite rare regarding the whole (five bifurcation unit) airway geometry model.

\begin{figure}
\centerline{\includegraphics{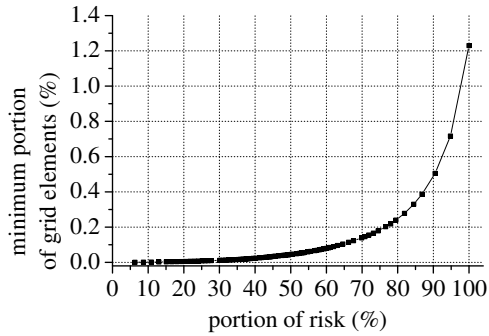}}
\caption{Minimum portion of grid elements contributing to a given portion of risk.}\label{fig:05}
\end{figure}

The spatial inhomogeneity of surviving fractions implies high spatial differences in the local probabilities of carcinogenesis. By the application of the model, the probability that at least one epithelial cell in the airway model takes both stochastic steps towards cancer can be computed. Let this probability be called risk. Computing the probabilities that at least one cell will be initiated and promoted in a given surface unit, the contribution to the risk of that grid element can be determined. As it is expected, the risk defined above originates only from a restricted area of the model geometry. Figure \ref{fig:05} demonstrates how large ratio of grid elements is responsible for a given proportion of risk at the 23 minutes exposure in the mine. As it is demonstrated, risk originates from hardly more than 1\% of all the grid elements and approximately 0.1\% of the surface of the airways provides 60\% of the whole risk computed by summing the result of relation~\ref{eq:04} calculated for the individual grid elements.

\begin{figure}
\centerline{\includegraphics{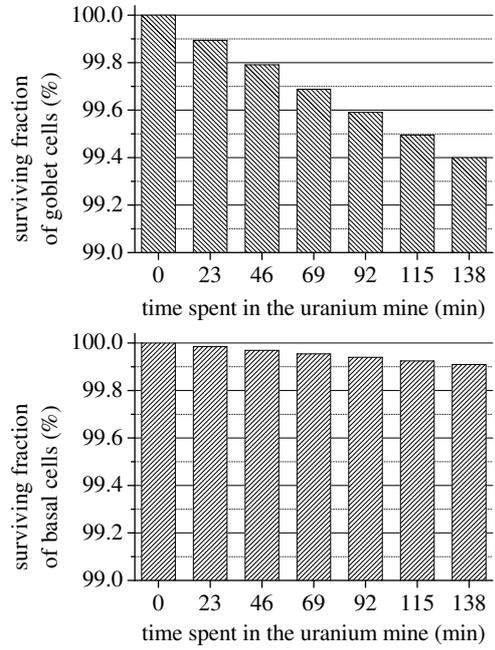}}
\caption{Surviving fraction as the function of radiation burden in case of goblet (upper panel) and basal cells (bottom panel) for the whole geometry.}\label{fig:06}
\end{figure}

Finally, some computations have been done to study, how the risk depends on the total burden. The number of deposited radon-progenies has been multiplied by whole numbers from 2 to 6 modeling the radiation burden during 46, 69, 92, 115 and 138 minutes of work in the New Mexico uranium mine. In this range, the surviving fractions decrease linearly by the macroscopic radiation burden. In case of goblet cells, obviously, this decrease is much higher, than in case of basal cells. The ratio of dead cells seems to be very small in both cases (0.60\% in case of goblet and 0.09\% in case of basal cells). However, if there were no radon-inhalation, only 0.13\% of all the cells would die during 138 minutes. This value is computed by the normal mitotic rate (1/30 $\mathrm{day}^{-1}$) and the stem cell ratio among all cells. The effect in cell killing, therefore, is quite significant, even in case of the whole geometry, data of which are plotted in Figure \ref{fig:06}.

\begin{figure}
\centerline{\includegraphics{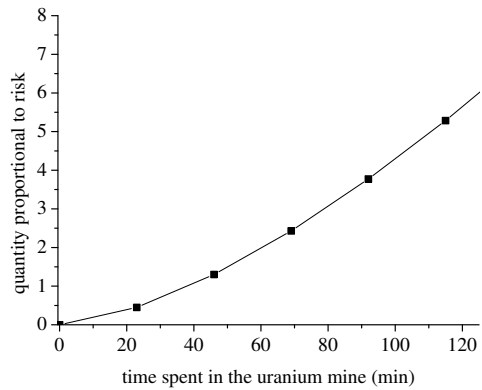}}
\caption{Quantity proportional to risk, i.e. at least one cell in the model airway geometry will be initiated and promoted as the function of radiation burden.}\label{fig:07}
\end{figure}

The exact value of risk cannot be determined because of the unknown proportion factor in relation (1), however the shape of the risk curve can be presented as the function of macroscopic radiation burden. Figure \ref{fig:07} demonstrates that risk in the modeled airways starts to increase non-linearly. This relationship differs from the functions applied for initiation, promotion and survival~(see relations 1-3), because tissue response to radiation is the superposition of these effects.

\section{Conclusions}

In case of radon inhalation, the cellular hit and absorbed dose distributions are very inhomogeneous within the central airways. There is three orders of magnitude difference between cellular and tissue doses in case of a 23-minute-long radon exposure in the mine, what means that the spatial heterogeneities should not be neglected in risk-assessment. The spatial distribution of surviving fractions in grid elements proves that not only single cells, but also cell clusters receive high radiation burdens. Therefore, in some parts of the central airways very high cell death frequencies can be observed even in case of the applied 23-minute-long exposure in the New Mexico uranium mine corresponding to 0.0129~WLM. These very high cell death frequencies supports the idea that processes governing lung cancer formation induced by radon-daughters may differ significantly from carcinogenesis caused by other types of exposure. In accordance with histological studies \cite{Sch01}, the results demonstrate that cancer formation probability is much higher in some regions of the central airways than in other parts of it. Importantly, risk seems to depend non-linearly on macroscopic radiation dose differing from the functions applied for initiation and promotion, which represent a radiation effect on individual cells and a radiation effect interpretable only on tissue fragments (interacting cells). This simple model demonstrates, that the interaction between radiation effects on different organizational levels may strongly influence the radiation response of the organism. Noteworthy, important effects such as dose-rate-effect and non-targeted effects are not considered in this study, therefore the computed values of biological quantities are quite uncertain.


\begin{thebibliography}{10.}

\bibitem{EPA01} United States Environmental Protection Agency (2003) Assessment of risks from radon in homes, {\it EPA Publications}, 402-R-03-003.
\bibitem{Bal01} Balásházy, I., Farkas, Á., Madas, B. G. and Hofmann,W. (2009) Non-linear relationship of cell hit and transformation probabilities in a low dose of inhaled radon progenies, {\it J Radiol Prot}, {\bf 29}, 147-162.
\bibitem{Far01} Farkas, Á. and Balásházy, I. (2008) Quantification of particle deposition in asymmetrical tra\-che\-obron\-chial model geometry, {\it Comp Biol Med}, {\bf 38}, 508-518.
\bibitem{Sch01} Schlesinger, R. B. and Lippmann, M. (1978) Selective particle deposition and bronchogenic carcinoma, {\it Environ Res}, {\bf 15}, 424-431.
\bibitem{Mer01} Mercer, R. R., Russell, M. L., Roggli, V. L. and Crapo,~J.~D. (1994) Cell number and distribution in human and rat lungs, {\it Am J Resp Cell Mol Biol}, {\bf 10}, 613-624.
\bibitem{Mer02} Mercer, R. R., Russell, M. L. and Crapo, J. D. (1991) Radon dosimetry based on the depth distribution of nuclei in human and rat lungs, {\it Health Phys}, {\bf 6}, 117-130.
\bibitem{SRI01} Ziegler, J. F., Biersack, J. P. and Ziegler, M. D. (2008) SRIM - The stopping and range of ions in matter, {\it Ion Technology Press}, The authors used the SRIM-2008.3, available on http://www.srim.org/.
\bibitem{Szo01} Szőke, I., Farkas, Á., Balásházy, I. and Hofmann, W. (2009) Stochastic aspects of primary cellular consequences of radon inhalation, {\it Radiat Res}, {\bf 171}, 96-106.
\bibitem{Tru01} Truta-Popa, L. A., Hofmann, W., Fakir, H. and Cosma, C. (2008) Biology based lung cancer model for chronic low radon exposures, {\it AIP Conf Proc}, {\bf 1034}, 78-85
\bibitem{ICR01} International Commission on Radiological Protection (1994) Human Respiratory Tract Model for Radiological Protection. ICRP Publication 66, {\it Ann ICRP}, {\bf 24}
\end{thebibliography}
\end{document}